\documentclass[a4paper]{jpconf}
\usepackage{graphicx}
\begin{document}
\title{Baryon and Lepton Numbers: Life in the Desert\footnote{Plenary talk given at the 18th International Symposium on Particles, Strings and Cosmology(PASCOS), Merida, Mexico, June 3-8, 2012. 
\\
The slides can be found in http://www.mpi-hd.mpg.de/personalhomes/fileviez/PASCOS-2012-Slides.pdf}}
\author{Pavel Fileviez Perez$^{a,b}$}

\address{
$^{a}$Particle and Astro-Particle Physics Division
\\
Max Planck Institute for Nuclear Physics (MPIK) 
\\
Saupfercheckweg 1, 69117 Heidelberg, Germany 
\\
$^{b}$Center for Cosmology and Particle Physics (CCPP) 
\\
New York University, 4 Washington Place, NY 10003, USA}
\ead{fileviez@mpi-hd.mpg.de}
\begin{abstract}
The simplest theories where we can understand the origin of the baryon and lepton number violating interactions are discussed.
We discuss the desert hypothesis in particle physics and the different scenarios where there is no need to assume it.
It is shown that the minimal supersymmetric B-L theory predicts lepton number violation at the Large Hadron Collider 
if supersymmetry is realized at the low scale. We present the BLMSSM where both symmetries, B and L, can be spontaneously 
broken at the TeV scale.
\end{abstract}
%
\section{Introduction}
%
It is well-known that in physics beyond the Standard Model (SM) very often one has to think about the violation of two SM symmetries, 
Baryon (B) and Lepton (L) numbers, and postulate the existence of a large desert between the low and high scales. 
It was pointed out by S. Weinberg~\cite{Weinberg:1979sa} long time ago that one can write down operators such as, 
${\cal O}_5 = c_\nu LL HH / \Lambda, \  \  \rm{and} \  \  {\cal O}_6 = c_{BL} QQQL /  \Lambda^2$,
where the first one breaks total lepton number and the second operator violates both symmetries. 
Typically, one can compute the coefficient in front of these operators in a grand unified theory defined 
at the high scale. The scale $\Lambda$ in ${\cal O}_6 $ has to be very large, i.e. $\Lambda  \ > \ 10 ^{14-16}$ GeV, 
in order to satisfy the bounds on the proton decay lifetime. See Fig.1 for a cartoon illustration of the desert 
hypothesis in particle physics.

If low energy supersymmetry is realized in nature the desert picture is different since at the TeV or 
multi-TeV scale we can have for example the Minimal Supersymmetric Standard Model (MSSM). 
In this case one can even predict the unification of gauge couplings at the unified scale, $M_{GUT} \approx 10^{16}$ GeV, 
if the desert hypothesis is true. See Fig.2 for a possible illustration of the desert picture in the case when 
we have Supersymmetry.  Unfortunately, in this case one has to face the fact that there are new interactions 
allowed by the SM gauge symmetry which violate B or L at the renormalizable level. 
These interactions are the well-known R-parity breaking interactions: $\hat{L} \hat{H}_u$, 
$\hat{Q} \hat{L} \hat{d}^c$, $\hat{L} \hat{L} \hat{e}^c$, and $\hat{u}^c \hat{d}^c \hat{d}^c$.   
%
\begin{figure}[t] 
\begin{center}
	\includegraphics[scale=0.35]{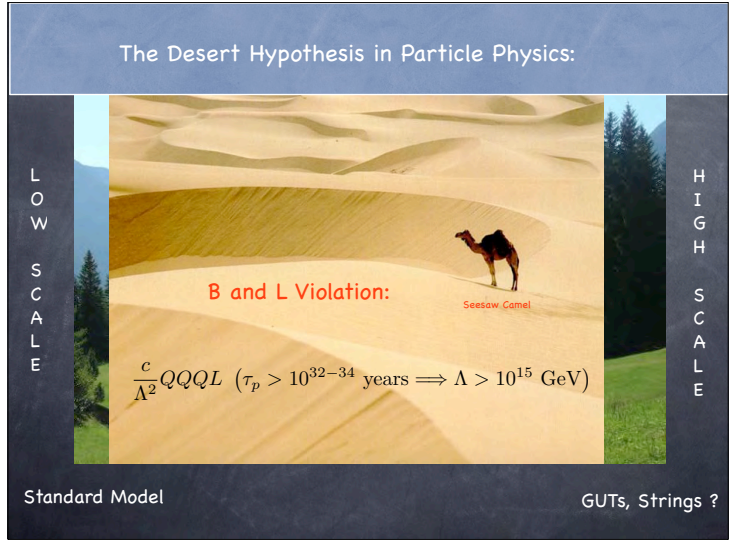}
\caption{Desert Hypothesis in Particle Physics. }
\end{center}	
\end{figure}
%

The main goal of this review is to discuss the possible origin of the B and L violating interactions in the Minimal Supersymmetric 
Standard Model and show the possibility to define a simple theory where one can have the spontaneous breaking of 
B and L  at the low scale. Therefore, in this case there is no need to postulate of the existence of a large desert. 
One has to say that the most of the people in the particle physics community got used to the idea of having a desert. 
In my opinion, it is hard to believe that there is no new physics between the low and high scales, and the fact that 
the people got used to it does not mean that this idea is true.  

What do we know about experimental evidences for B and L violation?
Unfortunately, after many experimental efforts still there is no direct evidences in a low energy process 
for the violation of baryon number. As it is well-known, many experimental collaborations have been looking 
for proton decay signals (a $\Delta B=1$ and $\Delta L=\rm{odd}$ process) and today the lower bounds on the proton decay lifetimes are very impressive. 
See Fig.3 for a summary of the proton decay lifetime bounds from different collaborations and Ref.\cite{review} for a review on proton decay.
\begin{figure}[b] 
\begin{center}
	\includegraphics[scale=0.35]{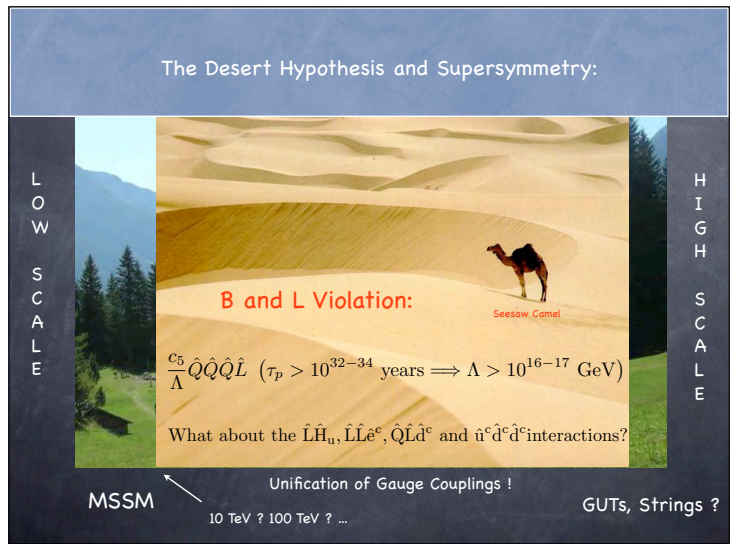}
	\caption{Supersymmetry and the Desert Hypothesis. }
\end{center}	
\label{img.NSSM.y}
\end{figure}
An interesting process where B must be broken in two units is $n-\bar{n}$ oscillation. 
In this case the bounds on the lifetime for free neutron oscillations are very weak, $\tau_{n-\bar{n}} > 10^8$ s~\cite{n-nbar}. 
There is a second $\Delta B =2$ process which can be used to set relevant constraints on new physics scenario, 
the di-nucleon decay $p \ p \to K^+ K^+$. Recently, a new bound has been found in Ref.~\cite{dinucleon} 
and the current lower bound is $\tau_{pp \to K^+ K^+} > 1.7 \times 10^{32}$ years. 
What about lepton number violation?. We know from neutrino oscillation experiments that the lepton number 
defined for each SM family is broken in nature, but still the total lepton number could be conserved. 
There are many experimental searches for neutrinoless double beta decay. In this case the 
total lepton number must be broken in two units and if it is discovered we can know about the 
Majorana nature of the neutrinos. For a review on neutrinoless double beta decay see Ref.~\cite{Rodejohann:2012xd}.
Then, what do we know from cosmology?. It is well-known that in order to explain the baryon asymmetry in the universe 
we need to have baryon number violation. Therefore, even if there is no a direct connection between the low energy 
processes mentioned above and the baryogenesis mechanism, we expect  baryon number violating processes in nature. 
 \begin{figure}[h] 
 \begin{center}
	\includegraphics[scale=0.35]{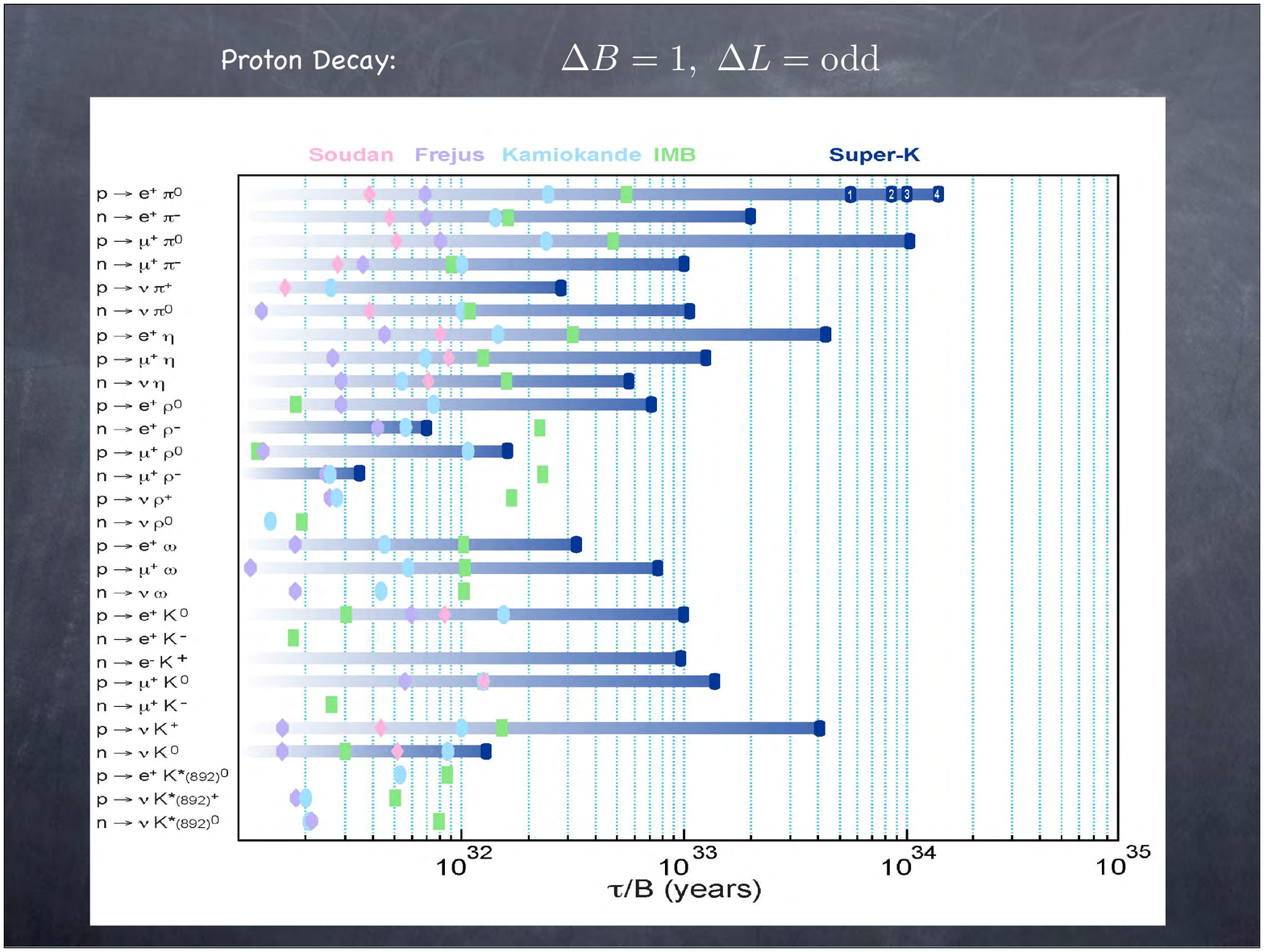}
	\caption{\small{Proton decay bounds~\cite{Miura}.}}
	\end{center}
\label{proton}
\end{figure} 
\section{B and L in the Superworld}
%
We have mentioned before that in the context of Supersymmetry we have interactions which break B and L at the 
renormalizable level and dimension five operators which give rise to proton decay:
\begin{eqnarray}
{\cal W}_{BL} &=& \epsilon \hat{L} \hat{H}_u \ + \  \lambda \hat{L} \hat{L} \hat{e}^c \ + \   \lambda^{'} \hat{Q} \hat{L} \hat{d}^c \ + \  \lambda^{''} \hat{u}^c \hat{d}^c \hat{d}^c \\
&+& \frac{c_\nu}{\Lambda} \hat{L} \hat{L} \hat{H}_u \hat{H}_u \ + \  \frac{c_L}{\Lambda} \hat{Q} \hat{Q} \hat{Q} \hat{L} \ + \  \frac{c_R}{\Lambda} \hat{u}^c \hat{d}^c \hat{u}^c \hat{e}^c.
\end{eqnarray}
The main difference between the terms in the first and second lines in this equation is that only the first interactions break matter parity. 
Matter parity is defined as $M=(-1)^{3(B-L)}$, being $-1$ for all matter chiral superfields and $+1$ for all other superfields.
Let us discuss some possible scenarios where these interactions could play a role. The possibility to generate 
Majorana neutrino masses  through the bilinear interaction, $\hat{L} \hat{H}_u$, has been studied by many groups.
See for example Ref.\cite{Valle} for a detailed analysis. A second scenario which give us very fast proton 
decay corresponds to the case when we combine the interactions $\hat{Q} \hat{L} \hat{d}^c$ and $\hat{u}^c \hat{d}^c \hat{d}^c$.
In this case one has the so-called dimension four contributions to nucleon decay. If the relevant coefficients are of order 
one and the sfermion masses are at the TeV scale the lifetime of the proton is too short, $\tau_p \sim 10^{-20}$ years. 
Therefore, one has two options: a) we assume small couplings, $\lambda^{'} \lambda^{''} < 10^{-26}$, or b) we try to understand 
the origin of these interactions to see if they are suppressed. It is important to mention that even if renormalizable interactions 
are absent still we get severe constraints on the dimension five operators mediating proton decay. 
See Ref.\cite{review} for a detailed discussion.

Now, let us focus on the first goal and try to understand the origin of the B and L violating interactions at the renormalizable level.
We have mentioned that these interactions break M-parity and of course B-L. Therefore, if we define a theory where 
B-L is conserved at the scale $\Lambda_{B-L}$ these interactions are not allowed before symmetry breaking and once we break 
the symmetry we can understand the size of these interactions. This idea has been studied by many different groups~\cite{R1,R2,R3,R4,R5,R6,R7,R8,R9,R10,R11,R12}.
We investigated this issue for the first in Ref.~\cite{R9} where we defined the simplest supersymmetric left-right model. In this case the only way to break 
the gauge symmetry $SU(2)_R \bigotimes U(1)_{B-L}$ to $U(1)_Y$ is to give a vacuum expectation value to the right-handed sneutrinos. Therefore, one can say that the minimal 
supersymmetric left-right model predicts that R-parity should be spontaneously broken and one expects lepton number violating signals at the Large Hadron Collider.
In this short review I will discuss mainly the simplest supersymmetric theory which predicts spontaneous breaking of R-parity and lepton number violation at the TeV scale.
%
\section{The Minimal Gauged $U(1)_{B-L}$ Model with Spontaneous R-parity Violation}
%
We have mentioned before that if we want to understand the possible origin of the lepton and baryon number violating interactions 
in the MSSM one has to define a theory where B-L is part of the gauge symmetry. Now, the simplest theory was proposed in 
Ref.\cite{R10} and here we discuss the main features and predictions: 
\begin{itemize}

\item This theory is based on the gauge group $SU(3)_C \bigotimes SU(2)_L \bigotimes U(1)_Y \bigotimes U(1)_{B-L}$.

\item The main prediction is that R-parity must be broken.

\item The only needed extra matter fields with B-L quantum numbers are the right-handed superfields.

\item The B-L and R-parity breaking scales are defined by the SUSY breaking scale.

\item  One predicts lepton number violation at the LHC.

\item The theory predicts two light sterile neutrinos below the eV scale.

\end{itemize}

The superpotential of this theory is very simple, one has the MSSM superpotential and an extra Yukawa interaction for the neutrinos,
\begin{equation}
{\cal W}_{B-L} = {\cal W}_{RpC} \ + \ Y_\nu \  \hat{L} \hat{H}_u \hat{\nu}^c,
\end{equation}
where
\begin{equation}
{\cal W}_{RpC} = Y_u \hat{Q} \hat{H}_u \hat{u}^c - Y_d \hat{Q} \hat{H}_d \hat{d}^c - Y_e \hat{L} \hat{H}_d \hat{e}^c.
\end{equation}
The new soft terms relevant for our discussion are given by
\begin{equation}
V_{soft} \supset  M_{\tilde{\nu}^c}^2 |\tilde{\nu}^c|^2 + \left(  A_\nu \tilde{L} H_u \tilde{\nu}^c \ + \  \rm{h.c.} \right).
\end{equation}
Now, in order to break the extra gauge symmetry, B-L, there is only one solution: {\it{The right-handed sneutrino must get a vacuum expectation value !}} 

It is very easy to show how the right-handed sneutrino gets a VEV. We compute the scalar potential and from the minimization conditions one finds
\begin{equation}
v_R^2 = - 8 \frac{M_{\tilde{\nu}^c}^2}{g_{BL}^2},
\end{equation} 
where $M_{\tilde{\nu}^c}^2$ has to be negative. It has been shown in Ref.~\cite{R11} that it is very easy to have a realistic spectrum 
for all sfermions in the theory. As one can expect the sfermion masses have a new contribution from the B-L D-term and 
typically the left-handed sleptons should be lighter if the soft mass for all the particles are similar at the TeV scale.

The neutrino spectrum in this theory is quite peculiar, it is easy to prove that only one of the right-handed neutrinos 
get a large mass around TeV scale and the rest are light. Therefore, one can have a 3 + 2 system with masses below the eV scale.
See Refs.~\cite{R3,3+2,Layers} for the detailed analysis. One can think about the possible new contributions coming 
from proton decay in this theory. Using the lepton number violating interactions above and the terms coming 
from the higher-dimensional operator $\hat{u}^c \hat{d}^c \hat{d}^c \hat{\nu}^c / \Lambda_B$ one finds new contributions 
to proton decay~\cite{R11} but assuming a large cutoff, around the GUT scale, one can satisfy the experimental bounds on 
the proton decay lifetime. Notice that here one needs to assume again a large desert from the TeV scale to the GUT scale.

This theory makes very interesting predictions for the LHC because the B-L symmetry is broken at the TeV scale 
and one has lepton number violating signals. Therefore, in order to test this theory one needs to look 
for the B-L gauge boson, identify the signals from the supersymmetric particles but in this case there 
is no missing energy signals associated to a stable LSP because R-parity is broken. 
We have investigated the most striking signals from this theory in Ref.~\cite{R11} where we 
have pointed out  the properties of the channels with four leptons, three of them with the same electric charge, 
and four jets. See Fig.4 for the topology of these signals and Ref.~\cite{R11} for the detailed analysis of these signals at the LHC.
 \begin{figure}[h] 
 \begin{center}
	\includegraphics[scale=0.35]{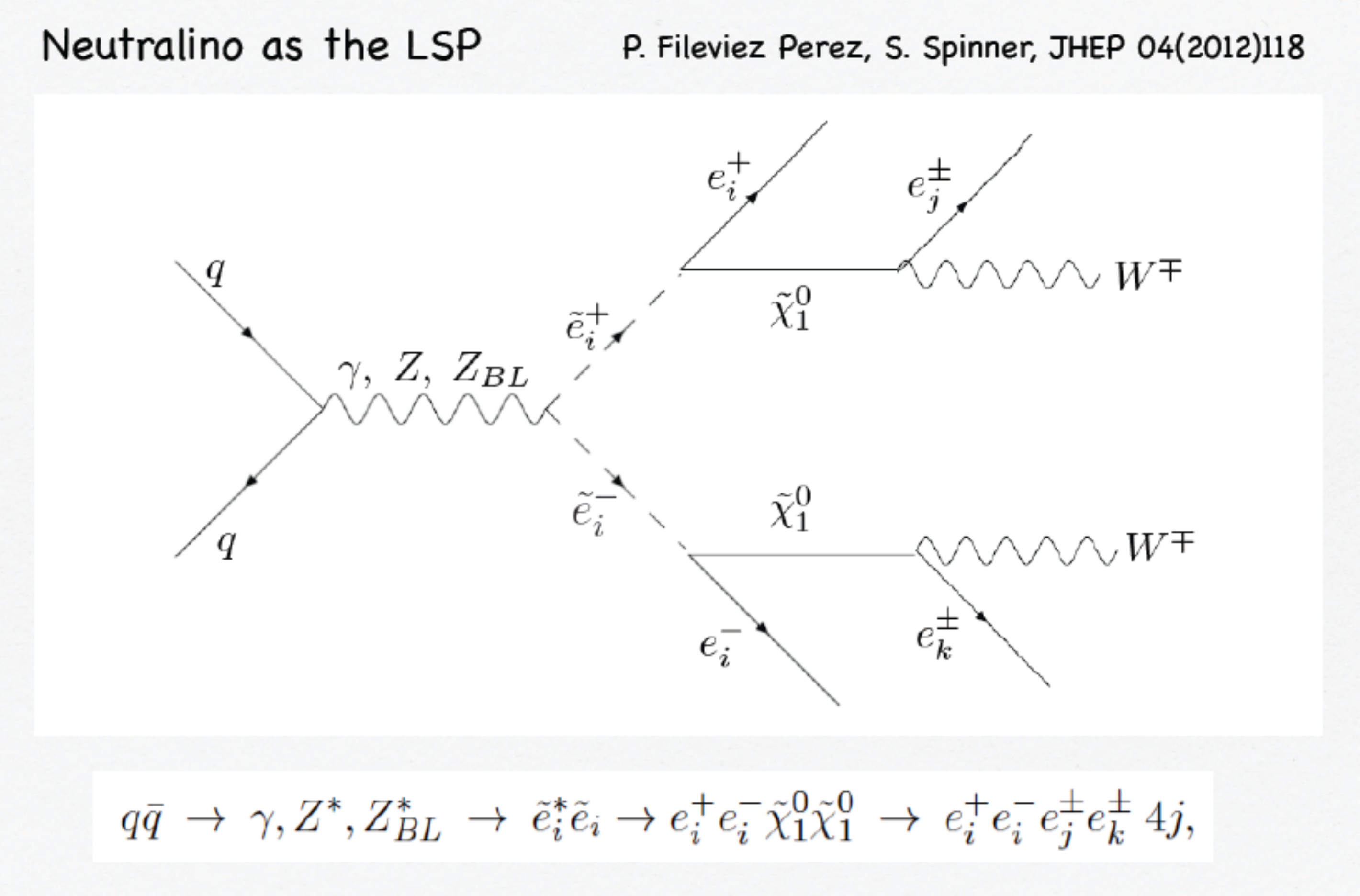}
	\caption{\small{Lepton Number Violating Signals~\cite{R11}.}}
	\end{center}
\label{signals}
\end{figure} 
Before we finish this section let us mention the main idea behind the simplest theory for R-parity conservation. 
In order to write down a simple theory where local B-L is broken to matter parity, we need to go beyond the 
minimal model discussed above and add extra Higgses. Since the right-handed neutrinos are present in the 
theory we always have the vacuum which corresponds to their vacuum expectation value. Therefore, 
we need to assume that only the extra Higgses with an even number of B-L break the gauge symmetry.
See also Ref.~\cite{Feldman,Nath} for the application of the Stueckelberg mechanism in this context.
Unfortunately, in the context of grand unified theories we cannot find a simple scenario for R-parity conservation.
It is well-known that in SUSY $SU(5)$ we need to impose matter parity by hand as in the MSSM, and in the context 
of $SO(10)$ we need huge representations to achieve R-parity conservation. Therefore, one can say that there 
is no a simple scenario based on GUTs for R-parity conservation.  Here, we would like to emphasize again that 
the minimal supersymmetric theory based on B-L makes a hard prediction: R-parity must be spontaneously broken 
and one expects striking lepton number violating signals at the LHC if supersymmetry is realized at the low scale. 
%
\section{Life in the Desert: Breaking B and L at the TeV scale}
%
We have discussed the possibility to understand the origin of the lepton and baryon number 
violating interactions present in the MSSM at the renormalizable level. Unfortunately, 
in this case one needs to assume a large cutoff in order to satisfy the bounds on the proton decay lifetime.
Here we will discuss a different theory where B and L can be broken at the TeV scale and there 
is no need to assume a desert. 

In order to prove that B and L can be broken at the low scale we can define a theory where we have two 
extra local symmetries. Then, the theory is based on $G_{SM} \bigotimes U(1)_B \bigotimes U(1)_L$\cite{BL} 
where $G_{SM}$ is the gauge group of the Standard Model. We refer to this model as ``BLMSSM". 
The main features of this theory are:

\begin{itemize}

\item The baryon and lepton numbers are local symmetries broken at the SUSY scale.

\item In order to cancel all anomalies we add a vector-like generation which is composed of 
$\hat{Q}_4$, $\hat{u}^c_4$, $\hat{d}^c_4$, $\hat{L}_4$, $\hat{e}^c_4$, $\hat{\nu}^c_4$, 
and $\hat{Q}_5^c$, $\hat{u}_5$, $\hat{d}_5$, $\hat{L}_5^c$, $\hat{e}_5$, $\hat{\nu}_5$.
Here the baryon and lepton numbers of the new chiral superfields are different since one 
has to satisfy the conditions $B_{Q_4} + B_{Q^c_4}=-1$ and $\ell_{L_4} + \ell_{L^c_4}=-3$.

\item In order to break the local baryonic symmetry and generate vector-like masses for the extra quarks we need to add new Higgses.
In this case we need to have for example the term $\hat{Q}_4 \hat{Q}^c_4 \hat{S}_B$. Therefore, this term defines 
the baryonic quantum number of $\hat{S}_B \sim 1$, and we have $\hat{\bar{S}}_B \sim -1$.  As we pointed out in Ref.~\cite{BL} one cannot 
avoid the operator $\hat{u}^c \hat{d}^c \hat{d}^c \hat{S}_B / \Lambda_B$, and after symmetry breaking one has baryon number violating 
interactions which can be suppressed by the cutoff of the theory.

\item In the leptonic sector the situation is more involved since one can have different scenarios:

a) In the first case if one sticks to the seesaw mechanism the new Higgses should have an even leptonic number, $\hat{S}_L, \hat{\bar{S}}_L \sim \pm 2$.
The new leptons have only chiral masses and there is no proton decay after symmetry breaking because all the interactions violate 
the total lepton number in an even number. This case was investigated in details in Ref.~\cite{BL} and Ref.~\cite{Arnold}.

b) One can generate vector-like masses for the new leptons, $\hat{L}_5^c \hat{L}_4 \hat{S}_L$, if the Higgses $\hat{S}_L, \hat{\bar{S}}_L \sim \pm 3$. 
In this case one gets operators mediating proton decay since $|\Delta B|=1$ and $|\Delta L|=3$. 
One of the operators that mediate proton decay in this case has dimension eleven, $\hat{Q} \hat{Q} \hat{Q} \hat{L} (\hat{L} \hat{H}_u)^2 \hat{S}_B \hat{\bar{S}}_L/ \Lambda^7$.    
Therefore, the cutoff scale in this case does not need to be very large.

\item There is no extra flavour violation at tree level because the new fermions do not mix with the SM fermions. 

\item There is no Landau pole at the low scale.
 
\end{itemize}

It is important to mention that in the context of the BLMSSM one could modify the LHC bounds on the supersymmetric particles 
because one has baryon number violation at the low scale. We have investigated in detail the impact of the new fermions 
on the Higgs mass~\cite{FileviezPerez:2012iw} and decays in Ref.~\cite{Arnold}. 
  
\section{Summary}

We have discussed in a general way the need to postulate a desert between the low and high scales in order to satisfy 
the bounds on the proton decay lifetime. In the first part of this review we have shown that the minimal supersymmetric 
B-L theory predicts that R-parity should be spontaneously broken and one expects lepton number violation at the LHC.
The simplest scenarios for the conservation of R-parity were briefly discussed. It has been mentioned that 
there is no simple grand unified theory defined in four dimensions where we understand the conservation of R-parity.
A simple theory where there is no need for a desert has been discussed. In this case B and L can be broken at the low scale 
and we do not get any operator mediating proton decay. It is important to emphasize that the two scenarios discussed here 
can be realized at the TeV scale and one can have very interesting supersymmetric signals at the Large Hadron Collider.

\section{Acknowledgments}

I would like to thank the organizers of the 18th International Symposium on Particles, Strings and Cosmology (PASCOS2012) 
in Merida for the invitation and warm hospitality. It is a great pleasure to thank my collaborators J. M. Arnold, V. Barger, 
B. Fornal, P. Nath, S. Spinner, and M. B. Wise for many discussions and enjoyable collaborations. This work was 
partially supported by the James Arthur Fellowship, CCPP, New York University.

\section*{References}


\begin{thebibliography}{9}

\bibitem{Weinberg:1979sa}
  S.~Weinberg,
  ``Baryon and Lepton Nonconserving Processes,''
  Phys.\ Rev.\ Lett.\  {\bf 43} (1979) 1566.

\bibitem{Miura}
M. Miura, Talk given at the 2011 International Workshop on Baryon and Lepton Number Violation (BLV2011), UTK, USA, 2011.

\bibitem{review}
  P.~Nath and P.~Fileviez Perez,
  ``Proton stability in grand unified theories, in strings and in branes,''
  Phys.\ Rept.\  {\bf 441} (2007) 191.

\bibitem{n-nbar}
Y. Kamyshkov, Talk given at the Spontaneous Workshop VI, Cargese, May 11, 2012.

\bibitem{dinucleon}
Michael D. Litos, ``A search for dinucleon decay into kaons using
the SK water cherenkov detector", Ph.D. Thesis, Boston
University, 2010.

\bibitem{Rodejohann:2012xd}
  J.~Beringer {\it et al.}  [Particle Data Group Collaboration],
  ``Review of Particle Physics (RPP),''
  Phys.\ Rev.\ D {\bf 86} (2012) 010001.
  
\bibitem{Valle}
  M.~Hirsch, M.~A.~Diaz, W.~Porod, J.~C.~Romao and J.~W.~F.~Valle,
  ``Neutrino masses and mixings from supersymmetry with bilinear R parity violation: A Theory for solar and atmospheric neutrino oscillations,''
  Phys.\ Rev.\ D {\bf 62} (2000) 113008
   [Erratum-ibid.\ D {\bf 65} (2002) 119901]
  [hep-ph/0004115].

\bibitem{R1}
  C.~S.~Aulakh and R.~N.~Mohapatra,
  ``Neutrino as the Supersymmetric Partner of the Majoron,''
  Phys.\ Lett.\ B {\bf 119} (1982) 136.

\bibitem{R2}
  M.~J.~Hayashi and A.~Murayama,
  ``Radiative Breaking of $SU(2)_R \times U(1)_{B-L}$ gauge symmetry induced by broken $N=1$ Supergravity in a Left-Right symmetric model,''
  Phys.\ Lett.\ B {\bf 153} (1985) 251.
  
\bibitem{R3}
  R.~N.~Mohapatra,
  ``Mechanism For Understanding Small Neutrino Mass In Superstring Theories,''
  Phys.\ Rev.\ Lett.\  {\bf 56} (1986) 561.
  
\bibitem{R4}
  L.~M.~Krauss and F.~Wilczek,
  ``Discrete Gauge Symmetry in Continuum Theories,''
  Phys.\ Rev.\ Lett.\  {\bf 62} (1989) 1221.
  
\bibitem{R5}
  A.~Font, L.~E.~Ibanez and F.~Quevedo,
  ``Does Proton Stability Imply the Existence of an Extra Z0?,''
  Phys.\ Lett.\ B {\bf 228} (1989) 79.
  
\bibitem{R6}
  A.~Masiero and J.~W.~F.~Valle,
  ``A Model For Spontaneous R Parity Breaking,''
  Phys.\ Lett.\ B {\bf 251} (1990) 273.

\bibitem{R7}
  S.~P.~Martin,
  ``Some simple criteria for gauged R-parity,''
  Phys.\ Rev.\ D {\bf 46} (1992) 2769
  [hep-ph/9207218].

\bibitem{R8}
  C.~S.~Aulakh, A.~Melfo, A.~Rasin and G.~Senjanovic,
  ``Seesaw and supersymmetry or exact R-parity,''
  Phys.\ Lett.\ B {\bf 459} (1999) 557
  [hep-ph/9902409].

\bibitem{R9}
  P.~Fileviez Perez and S.~Spinner,
  ``Spontaneous R-Parity Breaking and Left-Right Symmetry,''
  Phys.\ Lett.\ B {\bf 673} (2009) 251
  [arXiv:0811.3424 [hep-ph]].

\bibitem{R10}
  V.~Barger, P.~Fileviez Perez and S.~Spinner,
  ``Minimal gauged U(1)(B-L) model with spontaneous R-parity violation,''
  Phys.\ Rev.\ Lett.\  {\bf 102} (2009) 181802
  [arXiv:0812.3661 [hep-ph]].

\bibitem{R11}
  P.~Fileviez Perez and S.~Spinner,
  ``The Minimal Theory for R-parity Violation at the LHC,''
  JHEP {\bf 1204} (2012) 118
  [arXiv:1201.5923 [hep-ph]].

\bibitem{R12}
  P.~Fileviez Perez and S.~Spinner,
  ``The Fate of R-Parity,''
  Phys.\ Rev.\ D {\bf 83} (2011) 035004
  [arXiv:1005.4930 [hep-ph]].
  
\bibitem{3+2}
  D.~K.~Ghosh, G.~Senjanovic and Y.~Zhang,
  ``Naturally Light Sterile Neutrinos from Theory of R-parity,''
  Phys.\ Lett.\ B {\bf 698} (2011) 420
  [arXiv:1010.3968 [hep-ph]].
   
\bibitem{Layers}
  V.~Barger, P.~Fileviez Perez and S.~Spinner,
  ``Three Layers of Neutrinos,''
  Phys.\ Lett.\ B {\bf 696} (2011) 509
  [arXiv:1010.4023 [hep-ph]].
  
\bibitem{Feldman}
  D.~Feldman, P.~Fileviez Perez and P.~Nath,
  ``R-parity Conservation via the Stueckelberg Mechanism: LHC and Dark Matter Signals,''
  JHEP {\bf 1201} (2012) 038
  [arXiv:1109.2901 [hep-ph]].
  
\bibitem{Nath}
  P.~Nath,
  ``SUGRA Grand Unification, LHC and Dark Matter,''
  arXiv:1207.5501 [hep-ph].
     
\bibitem{BL}
  P.~Fileviez Perez and M.~B.~Wise,
  ``Breaking Local Baryon and Lepton Number at the TeV Scale,''
  JHEP {\bf 1108} (2011) 068
  [arXiv:1106.0343 [hep-ph]].
  
\bibitem{Arnold}
  J.~M.~Arnold, P.~Fileviez Perez, B.~Fornal and S.~Spinner,
  ``On Higgs Decays, Baryon Number Violation, and SUSY at the LHC,''
  Phys.\ Rev.\ D {\bf 85} (2012) 115024
  [arXiv:1204.4458 [hep-ph]].
  
\bibitem{FileviezPerez:2012iw}
  P.~Fileviez Perez,
  ``SUSY Spectrum and the Higgs Mass in the BLMSSM,''
  Phys.\ Lett.\ B {\bf 711} (2012) 353
  [arXiv:1201.1501 [hep-ph]].
  

\end{thebibliography}
\end{document}